\documentclass{PoS_mod}
\usepackage[english]{babel}
\usepackage[utf8]{inputenc}
\usepackage{amsmath}
\usepackage{amsfonts}
\usepackage{amssymb}
\usepackage{mathtools}
\usepackage{csquotes}
\usepackage{graphicx}
\usepackage{scalerel,stackengine}
\usepackage[backend=bibtex]{biblatex}

\bibliography{ICRC.bib}

\graphicspath{ICRC_plots/}

\title{Towards a fast and precise forward model for air shower radio simulation}

\ShortTitle{Towards a fast and precise forward model for air shower radio simulation}

\author{\firstauth{David Butler}\\
        IKP, Karlsruher Institut fur Technologie, Postfach 3640, 76021 Karlsruhe, Germany\\
        E-mail: \email{david.butler@kit.edu}}

\author{\speaker{Tim Huege}\\
        IKP, Karlsruher Institut fur Technologie, Postfach 3640, 76021 Karlsruhe, Germany\\
        E-mail: \email{tim.huege@kit.edu}}

\author{Olaf Scholten\\
        KVI - Center for Advanced Radiation Technology, Zernikelaan25 NL-9747 AA Groningen, The Netherlands\\
        E-mail: \email{scholten@kvi.nl}}

\abstract{The radio detection method for cosmic rays relies on coherent emission from electrons and positrons which is beamed in a narrow cone along the shower axis. Currently the only models to reproduce this emission with sufficient accuracy are Monte Carlo based simulations of the particle and radio emission physics, which require large investments of computation time. 

The work presented here focuses on condensing the simulation results into a semi-analytical model. This relies on building a framework based on theoretical predictions of radio emission, but instead of calculating the radio signal directly these models are used to map template simulations to the specifications of a given radio event. Our current approach slices the radio signal based on atmospheric depth of origin and weights these slices based on a shower parameter such as electron number or an effective dipole moment. One significant gain over the existing Monte Carlo codes lies in the fact this makes the depth of the shower maximum a direct input to the simulation where currently one has to pre-select showers based on their random number seed.

Such a model has great potential for heavily simulation-based analysis methods, for example the LOFAR air shower reconstruction. These techniques are severely limited by the available computation time but have the lowest errors in real measurement applications.}

\FullConference{35th International Cosmic Ray Conference\\
		 10-20 July, 2017\\
		 Bexco, Busan, Korea}

\begin{document}


\section{Introduction}

Current state-of-the-art radio emission models for cosmic ray air showers rely on simulating the particle microphysics. While this approach encompasses all our knowledge of the underlying processes and thus represents the best achievable accuracy it comes at a hefty cost: the Monte Carlo simulations \cite{Heck:1998vt}\cite{AlvarezMuniz:2011bs}\cite{Huege:2013vt} consume large amounts of computation time and their probabilistic nature means most analyses require many individual cascades to be computed in order to gauge statistical fluctuations. Due to measurement limitations current experiments often fall back to empirically derived analytical metrics extracted from the full simulations as is done by AERA \cite{Ewa:icrc}\cite{Aab:2015vta} and Tunka-Rex \cite{Kostunin:2015taa}. These methods pose additional limitations on reconstruction accuracy which is why other techniques compare measurements to a bank of simulations \cite{Buitink:2014eqa}.

On the other hand there are the macroscopic analytical models aiming to predict the radio signal through classical electrodynamics \cite{deVries:2010pp}\cite{Scholten:2016bpw}. They require far less computation but have failed to reproduce more recent measurements with sufficient accuracy to see widespread use.

We intend to combine these two approaches, building a hybrid scheme to analytically recombine templates extracted from CoREAS simulations \cite{Huege:2013vt}.

\section{Base Model}

The basis for our approach is the ability of CORSIKA to aggregate individual particles such as the histograms used in REAS \cite{Ludwig:2010pf}. The development of these shower properties can then be compared to the radio signal obtained from the contributing particles. 

A large amount of the total phase space describing the radio signal is tied to the physical locations of the receiving antennas. Real experiments only feature a limited number of antenna positions and this is reflected in the endpoint formalism \cite{James:2010vm}, which describes the electromagnetic field for a single line of sight. Since the position of the air shower relative to the experiment varies our model must be able to cover the full continuum of possible antenna positions nonetheless. Based on a limited number of template positions one can interpolate the radio signal given theoretical predictions of the two main emission mechanisms: Geomagnetic emission has a homogeneous polarisation aligned with the $\vec{v}\times\vec{B}$ axis. Charge excess emission produces a radially symmetric polarisation pattern. The field strength should be radially symmetric for each process individually \cite{Glaser:2016qso}, known asymmetries in the lateral distribution would arise from the vector sum of the two components. In the lateral direction interpolation of time series from fixed template positions looks promising for sufficiently short interpolation distances.

In the simplest case we are studying the development of the particle cascade is approximated as a scalar variable, such as the total number of radio-emitting particles, graphed over a one-dimensional evolution parameter such as time or height. Based on existing work on cascade theory and shower universality \cite{Lafebre:2009en} the best evolution parameter is atmospheric slant depth, defined here as the line-of-sight integral over density along the travel direction of the primary particle $\vec{v}$:
\begin{equation}
X_{slant}(l) = \int_{\infty}^{l} \rho ~\mathrm{d}\vec{l}
\end{equation}
The measurable radio time series can be subdivided into contributions from different slant depth regions. Given a slice centre $X_j$, thickness $\Delta X$ and an antenna position $\vec{r}$ a slice time series can be defined as
\begin{equation}
\vec{E}(X_j, \vec{r}) = \sum_{i}^{N_\mathrm{particles}} \vec{E}_{i}(\vec{r})
\end{equation}
summing over all particles satisfying $X_j-\Delta X/2\leq X_\mathrm{slant}(i)<X_j+\Delta X/2$. If the slices form a true subdivision of the entire atmosphere range $[0;X_\mathrm{ground}[$ the measured time series can be recovered as the direct sum of the slice time series
\begin{equation}
\vec{E}_\mathrm{total}(\vec{r}) = \sum_{j}^{N_\mathrm{slices}} \vec{E}(X_j,\vec{r})
\end{equation}
This slicing might be insufficient for highly inclined showers as there will be a significant asymmetry in the ambient density above and below the shower axis. In addition to the disconnect between geometric distance and layers of atmospheric depth relevant for the classical propagation of radio signals the asymmetry may also be significant in the development of the particle cascade. This would fundamentally violate our assumption of radial symmetry mandating explicit correction.

\section{Template Synthesis}\label{sec:Nsynth}

Given a normalisation curve $N_{t\mathrm{(emplate)}}(X_j)$ and a list of radio time series $\vec{E}_{t}(X_j, \vec{r})$ as detected at a fixed ground position $\vec{r}$ the measurable total time series of a \textit{different} particle shower should be recovered by weighting each slice with that cascade's normalisation $N_{r\mathrm{(eal)}}(X_j)$:
\begin{equation}\label{eq:Nsynth}
\vec{E}_{s\mathrm{(ynthesis)}}(\vec{r}) = \sum_{j}^{N_\mathrm{slices}} \frac{N_r(X_j)}{N_t(X_j)} \cdot \vec{E}_{t}(X_j, \vec{r})
\end{equation}
The best parameter to describe the shower evolution is still unclear. A natural quantity to describe the particle cascade would be the number of particles as embodied in the reconstruction of its maximum $X_\mathrm{max}$ in many analyses. 
In the context of radio emission we 
use the total number of electrons and positrons in our initial tests, counting those traversing an imaginary detector plane perpendicular to the shower axis.
\begin{figure}
\begin{center}
\includegraphics[width=0.75\textwidth]{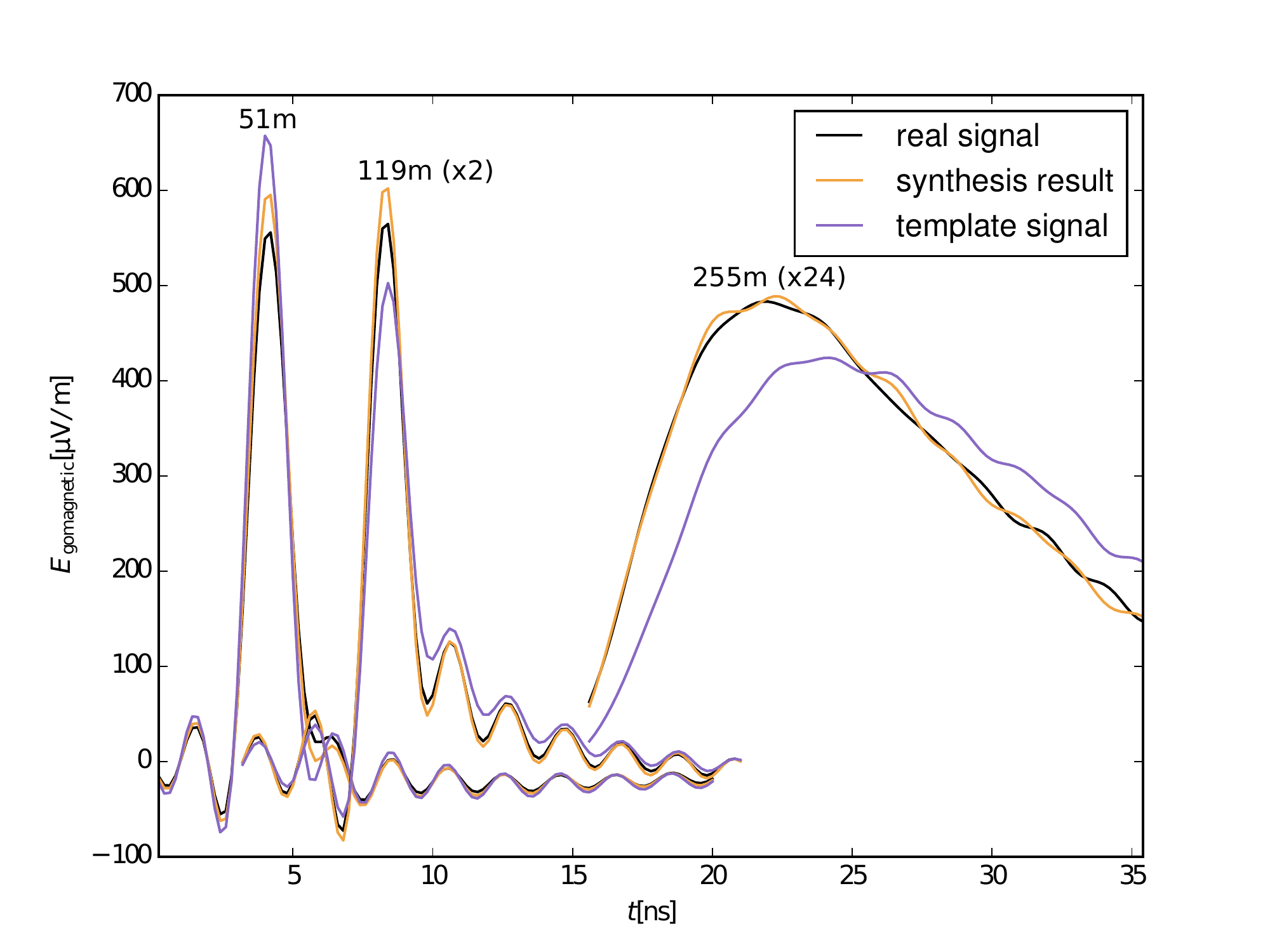}
\caption{Synthesised time series for different lateral distances. Both showers were simulated with CORSIKA7.5000 and the primary particles were $10^{17}\mathrm{eV}$ protons arriving vertically in both cases. The radio time series are filtered to $0-500\mathrm{MHz}$ in order to suppress incoherent emission and numerical noise which would not be observed by real experiments. All observer positions are located along the $\vec{v}\times\vec{v}\times\vec{B}$ axis. The black curves show the CoREAS output interpreted as the ``real'' signal, the blue curves are the template used in our model and the red curves display our synthesis result using equation \ref{eq:Nsynth}.}
\label{fig:Nsynth}
\end{center}
\end{figure}
This approach generally improves the similarity between the measurable radio time series of both showers as shown in figure \ref{fig:Nsynth}. For large lateral distances the renormalised time series of the synthesised pulse matches that of the real one almost perfectly. However for smaller lateral distances differences become apparent, and these distances show both lateral structure and shower-to-shower fluctuations. Thus a different normalisation or a more complex approach is necessary.

\section{Effective Dipole Construction}\label{sec:DP}

A simple particle count neglects the velocities and distances traveled by the particles which factor into the predicted radio signal. Further the particle number does not account for destructive interference due to opposite directions of travel. These factors could be accounted for through a different normalisation curve
\begin{equation}\label{eq:DP}
\vec{D}(X_j) = \sum_{i}^{N_\mathrm{particles}} q_{i} \cdot (\vec{d}_\mathrm{end}(i)-\vec{d}_\mathrm{start}(i))
\end{equation}
where $q_i$ is the particle charge, $\vec{d}_s(i)$ is the startpoint of the track, $\vec{d}_e(i)$ its endpoint and the sum is over all particle tracks $i$ which fulfill $X_j-\Delta X/2\leq X_\mathrm{slant}(i)<X_j+\Delta X/2$. Using the slant depth of the track starting point this is equivalent to our slicing of the electric field contribution ensuring physical consistency.

This vector-valued quantity can be interpreted as a local effective dipole moment ascribed to a point source in the centre of the slice. Based on theoretical descriptions of the emission mechanisms the bulk of the detected radio signal originates from interactions with the Earth's ambient magnetic field $\vec{B}$. This geomagnetic emission process \cite{Scholten:2007ky} should be represented by the $\vec{v}\times\vec{B}$-component of the effective dipole. Meanwhile the charge excess/Askaryan emission \cite{Askaryan:1962hbi} should be most closely correlated with the component parallel to the shower axis. In observations of atmospheric electric fields during thunderstorms their additional deflection of particles would also be visible in the dipole even if their number does not change significantly. 

\begin{figure}
\begin{center}
\includegraphics[width=0.7\textwidth]{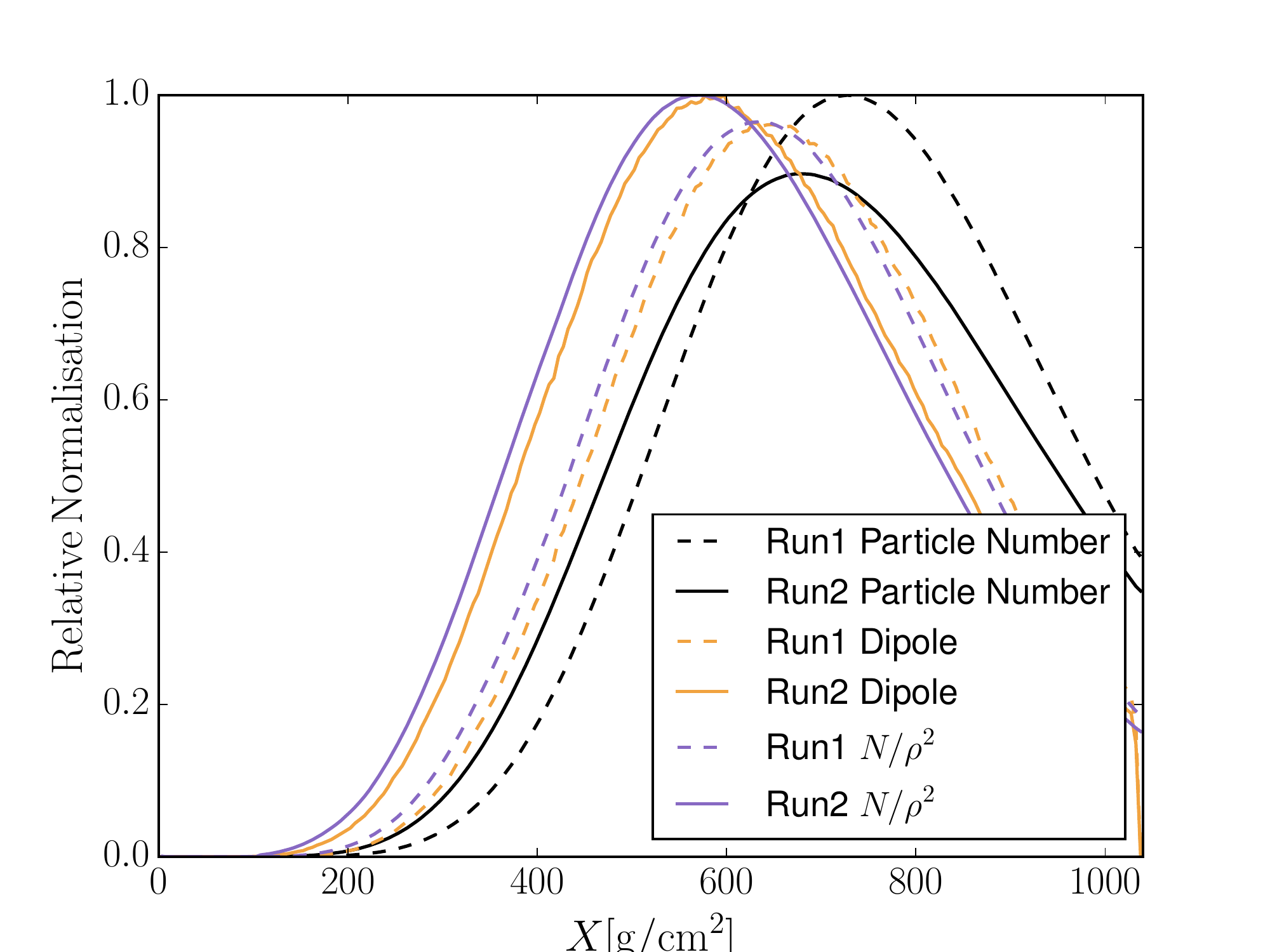}
\caption{Normalisation curves for the two showers used in our first synthesis shown in figure \ref{fig:Nsynth}, where run 1 was used as the template and run 2 as the ``real'' signal. All curves are scaled to the maximum value for that quantity between \textit{both} runs.}
\label{fig:DPnorm}
\end{center}
\end{figure}

Curiously the dipole $\vec{v}\times\vec{B}$-component is almost identical to the particle count performed by CORSIKA divided by the ambient density squared as shown in figure \ref{fig:DPnorm}, save for the global scale which is not relevant in our model. One factor of density originates from a difference in normalisation: the dipole is constructed as a volume integral while the particle number is counted in a plane. Since the slices are defined through atmospheric depth their geometric volume scales as the inverse of density when neglecting its variation within each slice. The second power reflects the geometrical track length, representative of the mean free path in the medium, also scaling with the inverse of the local density.
In the above simplistic rescaling the dipole performs consistently worse than the particle number. This is somewhat to be expected as the track length factors in the ambient density through the mean free path while the atmospheric depth is an explicit line integral over said density. Thus the dipole might overcompensate for density effects in this application at the expense of properly describing the development of the particle cascade. 

\section{Relative Evolution Synthesis}

The results from sections \ref{sec:Nsynth} and \ref{sec:DP} suggest we should account for the shower development in a more direct fashion. Rather than comparing slices of equal absolute atmospheric depth one can measure the relative depth to the shower maximum $\Delta X=X-X_\mathrm{max}$ instead. This is functionally equivalent to the definition of the ``relative evolution stage'' \cite{Lafebre:2009en}
\begin{equation}
t = \frac{X-X_\mathrm{max}}{X_0}
\end{equation}
provided the radiation length $X_0$ is assumed constant across the entire atmosphere. 

In matching templates to target positions we now assume the source to have a constant \textit{angular} emission pattern. This means the lateral axis distance at ground level will now differ between the real signal and its template position. Implicitly we also assume the structure of the radio pulse not to change along a given line of sight which is consistent with our approximation of the entire slice as a point source.

\section{Line-Of-Sight Scaling}
\begin{figure}
\begin{center}
\includegraphics[width=0.75\textwidth]{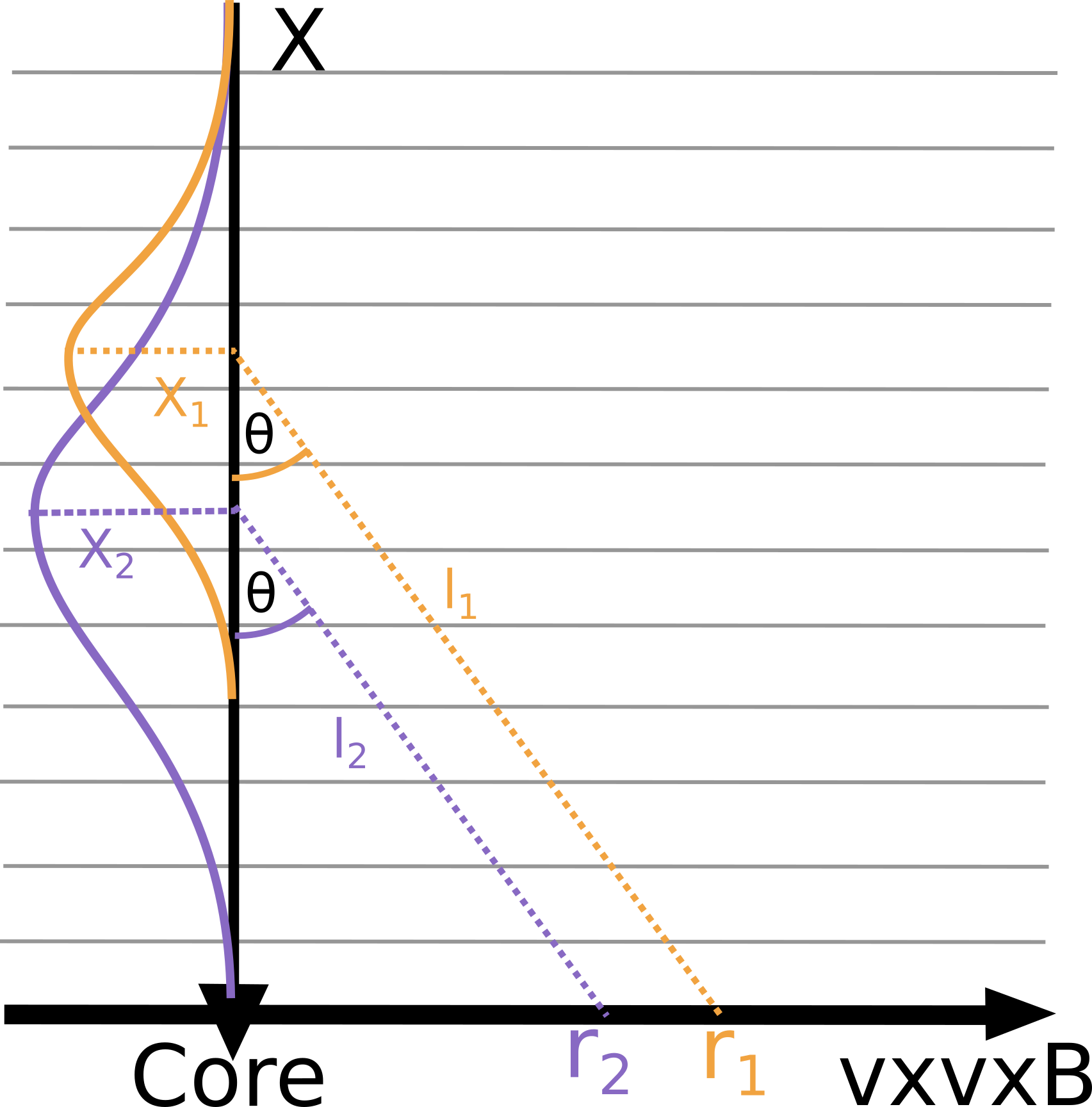}
\caption{Sketch of the geometry when comparing slices of different atmospheric depth for vertical air showers.}
\label{fig:deltaXsketch}
\end{center}
\end{figure}

\begin{figure}
\begin{center}
\includegraphics[width=0.75\textwidth]{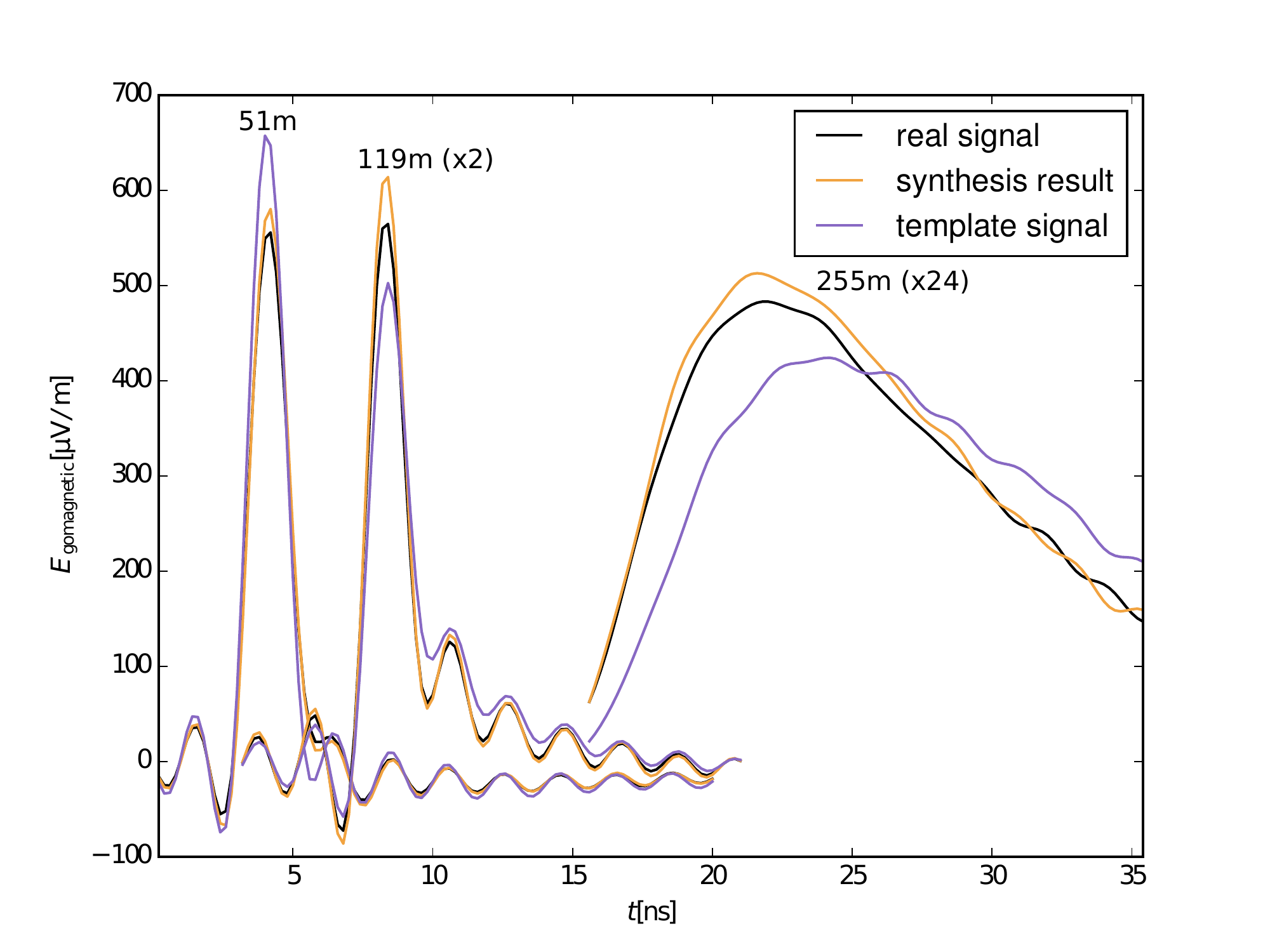}
\caption{Dipole-based (see equation \ref{eq:DP}) relative evolution synthesis test analogous to figure \ref{fig:Nsynth}. The particle cascades are identical, only the template antenna positions within CoREAS differ from the previous example. The blue template curves show the measurable signal of the template shower at the target positions, \textbf{not} the direct sum of the template time series used in the synthesis as the latter has no physical meaning.}
\label{fig:DPsynth}
\end{center}
\end{figure}

In order to compare slices of different absolute atmospheric depth as shown in figure \ref{fig:deltaXsketch} several other changes are necessary. The primary addition stems from the physical geometry of the system as the slices lie at different distances $l$ to the antenna. A radiative electric field is expected to decrease linearly with increasing distance, thus each slice must be rescaled accordingly:
\begin{equation}
\vec{E}_{1}(X_1, \vec{r}_1) = \vec{E}_{2}(X_2, \vec{r}_2) \cdot\frac{l_2(X_2, \vec{r}_2)}{l_1(X_1, \vec{r}_1)}
\end{equation}
Further we must compensate for the difference in arrival times. Our assumption here is that the particle cascade propagates at the vacuum speed of light along the shower axis while the radio signal travels from the slice centre to the antenna in a straight line with the speed of light in the medium $c/n$. The changing refractive index $n(\rho)$ is approximated through the same effective value calculation as in CoREAS, which assumes the local refractivity $\epsilon=n(\rho)-1$ scales linearly with the ambient density. Thus for two slices with $X_1<X_2$ and corresponding effective refractive indices $n_1$, $n_2$ averaged over their respective lines of sight $l_1$, $l_2$ the individual travel times are
\begin{equation}
t_1 = \frac{h(X_1)-h(X_2)}{c} + l_1\cdot\frac{n_{1}(\vec{l}_1)}{c} ~~~;~~~ t_2 = l_2\cdot\frac{n_{2}(\vec{l}_2)}{c}
\end{equation}
where $h$ is the distance along the shower axis. This yields a time delay of
\begin{equation}
\Delta t = t_1-t_2 = \frac{1}{c}\cdot\left(h(X_1)-h(X_2) + l_1\cdot n_{1}(\vec{l}_1)-l_2\cdot n_{2}(\vec{l}_2)\right)
\end{equation}
These additional steps are necessary for the relative evolution slicing approach to function while being completely absent from the simple example in section \ref{sec:Nsynth}. As shown in figure \ref{fig:DPsynth} the results currently are of a quality comparable to the performance of equation \ref{eq:Nsynth} but not better. The main disadvantage is the loss of consistent accuracy for large lateral distances, though the differences are still small in terms of peak height. As the line-of-sight scaling will also be relevant for non-vertical shower geometries we are still investigating both options.
\section{Conclusion}

We have outlined a new approach to combining the speed and elegance of analytical models with the accuracy of numerical simulations in describing the radio emission from cosmic ray air showers. With such a hybrid model modern analysis techniques for detailed air shower radio measurements could save significant amounts of computation time, or cover a much larger parameter space with equal investment.

Further one could envisage parameter-free inference methods for the development of the shower, such as a direct reconstruction of $X_\mathrm{max}$, which are currently unfeasible due to the computation demand and the ``black box'' nature of Monte Carlo simulations.

\printbibliography

\end{document}